\documentclass[12pt,thmsa,sw20aip]{article}
\usepackage[T1]{fontenc}
\usepackage{graphics}

\makeatletter

\newcommand{\LyX}{L\kern-.1667em\lower.25em\hbox{Y}\kern-.125emX\spacefactor1000}

\usepackage{psfrag,graphicx}

\makeatother

\begin{document}

\author{S.A. Artamonov\protect\( ^{a}\protect \), Yu.G. Pogorelov\protect\( ^{b}\protect \),
V.R. Shaginyan\protect\( ^{a}\protect \)\\
\protect\( ^{a}\protect \)\textit{Petersburg Nuclear Physics Institute, RAS,
Gatchina 188350, Russia};\\
 \protect\( ^{b}\protect \) \textit{Universidade do Porto, Centro de Física
de Porto, 4150 Porto, Portugal}}

\title{Ground state instability in systems of strongly interacting fermions}

\maketitle
\begin{abstract}
We analyze stability of a fermion system with model repulsive pair interaction
potential. The possibility for different types of restructuring of the Fermi
ground state (at sufficiently great coupling constant) is related to the analytic
properties of such potential. In particular, for the screened Coulomb law it
is shown that the restructuring cannot be of the Fermi condensation type, known
earlier for some exactly solvable models, and instead it belongs to the class
of topological transitions (TT). For this model, a phase diagram has been built
in the variables ''screening parameter - coupling constant'' which displays
two kinds of TT: a \( ^{5}/_{2} \)-kind similar to the known Lifshitz transitions
in metals, and a 2-kind characteristic for a uniform strongly interacting system. 
\end{abstract}

The common ground state of isotropic Fermi gas with density \( \rho  \) is
described (at zero temperature) by the stepwise Fermi function \( n_{{\textrm{F}}}(p)=\theta (p_{{\textrm{F}}}-p) \),
dropping discontinuously from 1 to 0 at the Fermi momentum \( p_{{\textrm{F}}}=(3\pi ^{2}\rho )^{1/3} \)
(in the units where \( \hbar =1 \)). The Landau theory of an interacting Fermi
liquid started from the assumptions that the quasiparticle distribution function
\( n(p) \) coincides with the one of an ideal gas, while the single particle
spectrum \( \varepsilon _{p} \) is similar to that of ideal gas, being characterized
by the effective mass \( M^{*} \) \cite{lan}. 

But these \( n_{{\textrm{F}}}(p) \) and \( \varepsilon _{p} \) can be broken
down under certain circumstances. The best known example is the Cooper pairing
at arbitrarily weak attractive interaction with subsequent formation of the
pair condensate and gapped quasiparticle spectrum \cite{bcs}, but also a repulsive
interaction, if sufficiently strong, can produce non-trivial ground states.
The first example of such restructuring for a Fermi system with model repulsive
interaction \cite{khosha} revealed existence of a critical value \( \alpha _{cr} \)
for the interaction constant \( \alpha  \), such that at \( \alpha =\alpha _{cr} \)
the stability criterion \( s(p)=(\varepsilon _{p}-\varepsilon _{{\textrm{F}}})/(p^{2}-p_{{\textrm{F}}}^{2})>0 \)
fails just at the Fermi surface: \( s(p_{{\textrm{F}}})=0 \) (\( p_{{\textrm{F}}} \)-instability).
Then at \( \alpha >\alpha _{cr} \) an exact solution exists of a variational
equation for \( n(p) \) (following from the Landau energy functional \( E[n(p)] \)),
exhibiting a certain finite interval \( (p_{1},p_{2}) \) around \( p_{{\textrm{F}}} \)
where the distribution function \( n(p) \) varies continuously and takes intermediate
values between \( 1 \) and \( 0 \), while the single-particle excitation spectrum
\( \varepsilon _{p} \) has a flat plateau, 
\begin{equation}
\label{flat}
\frac{\delta E[n(p)]}{\delta np)}=\varepsilon _{p}=\mu ;\; \; \; p_{1}\leq p\leq p_{2},
\end{equation}
 with \( \mu  \) being the chemical potential. It is seen from eq. (\ref{flat})
that the occupation numbers \( n(p) \) become variational parameters, deviating
from the Fermi function in order to minimize the energy \( E \). This phenomenon
was called fermion condensation (FC) and soon several model forms for \( E[n(p)] \)
were proposed \cite{vol}-\cite{ksz} providing similar solutions. However,
these models do not answer the question whether there exist other types of phase
transitions related to a rearrangement of the Fermi function \( n_{{\textrm{F}}}(p) \)
and the single particle spectrum \( \varepsilon _{p} \). It is pertinent to
note that the idea of multiconnected Fermi sphere, with production of new, interior
segments of the Fermi surface, has been considered already \cite{llvp,zb}.
This determines an interest to examine the overcritical regimes \( \alpha >\alpha _{cr} \)
for the models displaying alternative types of instability. 

The main goal of this Letter is to consider possible types of rearrangement
of the Fermi ground state. We shall relate the types of such rearrangement to
analytic properties of the single particle potential of a system in question
and show that there can exist phase transitions of different type from FC. 

The general scheme of Refs. \cite{khosha}-\cite{khveshch} for a homogeneous
system of non-polarized fermions with mass \( M \) and model isotropic interaction
potential \( U(p) \) considers the energy functional \( E[n(p)] \):

\begin{equation}
\label{energ}
E[n(p)]=\int \frac{p^{2}}{2M}n(p)\frac{d{\textbf {p}}}{(2\pi )^{3}}+\frac{1}{2}\int \int n(p)U(|{\textbf {p}}-{\textbf {p}}^{\prime }|)n(p^{\prime })\frac{d{\textbf {p}}d{\textbf {p}}^{\prime }}{(2\pi )^{6}},
\end{equation}
 and the related quasiparticle dispersion law: 
\begin{equation}
\label{disp}
\varepsilon _{p}=\frac{p^{2}}{2M}+\int U(|{\textbf {p}}-{\textbf {p}}^{\prime }|)n(p^{\prime })\frac{d{\textbf {p}}^{\prime }}{(2\pi )^{3}}.
\end{equation}
 Performing the angular integration and passing to the dimensionless variables:
\( x=p/p_{{\textrm{F}}} \), \( y=2M\varepsilon _{p}/p_{{\textrm{F}}}^{2} \),
\( z=2\pi ^{2}ME/p_{{\textrm{F}}}^{5} \), eqs. (\ref{energ},\ref{disp}) can
be presented in a simpler form: 
\begin{equation}
\label{zet}
z[\nu (x)]=\int [x^{4}+\frac{1}{2}x^{2}V(x)]\nu (x)dx,
\end{equation}

\begin{equation}
\label{disp1}
y(x)=x^{2}+V(x),
\end{equation}
 with the single particle potential \( V(x) \) being given by, 
\begin{eqnarray}
V(x) & = & \frac{1}{x}\int x^{\prime }\nu (x^{\prime })u(x,x^{\prime })dx^{\prime },\label{kern} \\
u(x,x^{\prime }) & = & \frac{M}{\pi ^{2}p_{F}}\int\limits _{|x-x^{\prime }|}^{x+x^{\prime }}u(t)tdt.\nonumber 
\end{eqnarray}
 Here \( u(x)\equiv U(p_{{\textrm{F}}}x) \), and the distribution function
\( \nu (x)\equiv n(p_{{\textrm{F}}}x) \) is positive, obeys the normalization
condition: 
\begin{equation}
\label{norm}
\int x^{2}\nu (x)dx=1/3,
\end{equation}
 and the Pauli principle limitation \( \nu (x)\leq 1 \). The latter can be
lifted using, e.g., the ansatz: \( \nu (x)= \) \( [1+\tanh \eta (x)]/2 \),
then the system ground state should correspond to the minimum of the functional:

\begin{equation}
\label{funct}
f[\eta (x)]=\int [1+\tanh \eta (x)]\{x^{4}-\mu x^{2}+\frac{1}{4}x\int x^{\prime }[1+\tanh \eta (x^{\prime })]u(x,x^{\prime })dx^{\prime }\}dx,
\end{equation}
 containing a Lagrange multiplier \( \mu  \), with respect to an \textit{arbitrary}
variation of the auxiliary function \( \eta (x) \). This permits to present
the necessary condition of extremum \( \delta f=0 \) as: 
\begin{equation}
\label{extr}
x^{2}\nu (x)[1-\nu (x)][y(x)-\mu ]=0,
\end{equation}
 which means that either \( \nu (x) \) takes only the values \( 0 \) and \( 1 \)
or the dispersion law is flat \cite{khosha}: \( y(x)=\mu  \), in accordance
with eq. (\ref{flat}). The last possibility just corresponds to FC. As it is
seen from eq. (\ref{flat}), the single particle spectrum \( \varepsilon _{p} \)
in this case cannot be analytic function of complex \( p \) in any open domain,
containing the FC interval \( \Delta p=[p_{1},p_{2}] \). In fact, all the derivatives
of \( \varepsilon _{p} \) with respect to \( p \) should be zero along \( \Delta p \),
while this is not the case along the real axis outside \( \Delta p \). For
instance, in the FC model with \( U(p)=U_{0}/p \) \cite{khosha} the kernel,
eq. (\ref{kern}), results non-analytic: 
\begin{equation}
u(x,x^{\prime })=\frac{MU_{0}}{\pi ^{2}p_{{\textrm{F}}}}(x+x^{\prime }+|x-x^{\prime }|)
\end{equation}
 which eventually causes non-analyticity of the potential \( V(x) \). 

On the other hand, as it follows from eq. (\ref{disp1}), the single particle
spectrum will be an analytic function along the whole real axis, provided \( V(x) \)
is such a function. In this case FC is forbidden and the only alternative to
the Fermi ground state (if the stability criterion gets broken) leaves in a
topological transition (TT) between the topologically unequal states with \( \nu (x)=0,1 \)
\cite{vol1}. Generally, all such states are classified by the indices of connectedness
(known as Betti numbers in algebraic topology \cite{naka}) for the support
of \( \nu (x) \). In fact, for an isotropic system, these numbers are simply
to count the separate (concentric) segments of the Fermi surface. Then the system
ground state will correspond to such a multiconnected distribution (Fig. 1):

\begin{equation}
\label{dist}
\nu (x)=\sum _{i=1}^{n}\theta (x-x_{2i-1})\theta (x_{2i}-x),\, 
\end{equation}
 that the parameters \( 0\leq x_{1}<x_{2}<\ldots <x_{2n} \) obey the normalization
condition:

\begin{equation}
\label{normm}
\sum _{i=1}^{n}(x_{2i}^{3}-x_{2i-1}^{3})=1,
\end{equation}
 and the related \( z \), eq. (\ref{zet}), 
\begin{equation}
\label{zetm}
z=\frac{1}{2}\sum _{i=1}^{n}\int\limits _{x_{2i-1}}^{x_{2i}}x^{2}[x^{2}+y(x)]dx
\end{equation}
 has the absolute minimum with respect to \( x_{1},\ldots ,x_{2n-1} \) and
to \( n\geq 1 \).\begin{figure}
\psfrag{v}[][][3]{\hspace{-5mm}\( \nu (x) \)}
\psfrag{1}[][][3]{\hspace{-3mm}1}
\psfrag{0}[][][3]{\hspace{-3mm}0}
\psfrag{a}[t][][3]{\( x_{1} \)}
\psfrag{b}[t][][3]{\( x_{2} \)}
\psfrag{c}[t][][3]{\( x_{3} \)}
\psfrag{d}[t][][3]{\( x_{4} \)}
\psfrag{e}[t][][3]{\( x_{2n-1} \)}
\psfrag{f}[t][][3]{\( x_{2n} \)}
\psfrag{h}[][][3]{\( \cdots  \) }

{\centering \resizebox*{10cm}{!}{\includegraphics{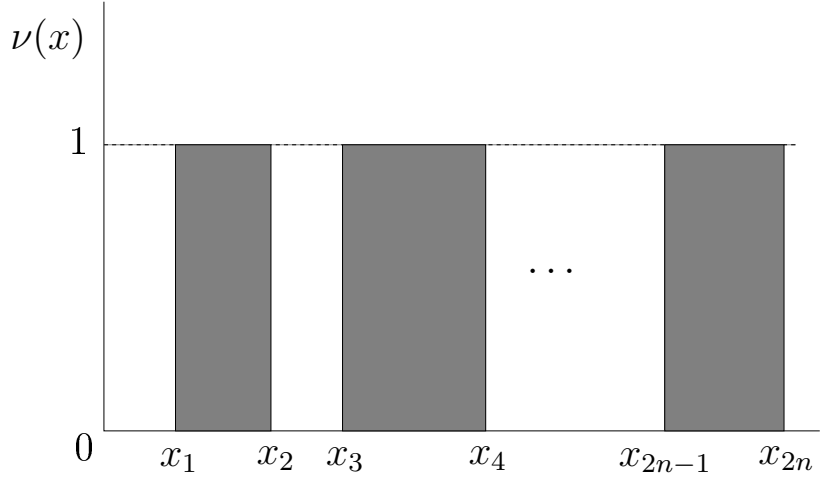}} \par}

\caption{\label{fig1}Occupation function for a multiconnected distribution.}
\end{figure} 

Differentiating eq. (\ref{zetm}) in the parameters \( x_{1},\ldots ,x_{2n-1} \)
with use of the relations: 
\begin{equation}
\label{deriv}
\frac{{\partial }x_{2n}}{{\partial }x_{k}}=(-1)^{k-1}(\frac{x_{k}}{x_{2n}})^{2},\, \, \, 1\leq k\leq 2n-1,
\end{equation}
 by virtue of eq. (\ref{normm}), and taking into account that the potential
\( V(x) \) in the dispersion law \( y(x) \) in fact also depends on these
parameters: 
\begin{equation}
\label{vm}
V(x)=\frac{1}{x}\sum _{i=1}^{n}\int _{x_{2i-1}}^{x_{2i}}x^{\prime }u(x,x^{\prime })dx^{\prime },
\end{equation}
 we present the necessary conditions of extremum as: 
\begin{equation}
\label{level}
\frac{{\partial z}}{{\partial }x_{k}}=(-1)^{k}x_{k}^{2}[y(x_{k})-y(x_{2n})]=0,\, \, \, 1\leq k\leq 2n-1.
\end{equation}
 Hence the multiconnected ground state is controlled by the evident rule of
\textit{unique Fermi level}: \( y(x_{k})=y(x_{2n}) \), for all \( 1\leq k\leq 2n-1 \)
(except for \( x_{1}=0 \)). In principle, given the kernel \( u(x,x^{\prime }) \)
all the \( (2n-1) \) unknown parameters \( x_{k} \) can be found from eqs.
(\ref{level},\ref{disp1},\ref{vm}). Then, the sufficient stability conditions:
\( \partial ^{2}z/\partial x_{i}\partial x_{j}=\alpha _{i}\delta _{ij} \),
\( \alpha _{i}>0 \), provide the generalized stability criterion: the dimensionless
function 
\begin{equation}
\label{stab}
\sigma (x)=2Ms(p)=\frac{y(x)-y\left( x_{2n}\right) }{x^{2}-x_{2n}^{2}}
\end{equation}
 should be \textit{positive within filled and negative within empty intervals,
turning zero at their boundaries} in accordance with eqs. (\ref{level}). It
can be proved rigorously that for given analytic \( u(x,x^{\prime }) \) this
criterion uniquely defines the system ground state. 

In what follows we shall label each multiconnected state, eq. (\ref{dist}),
by an entire number related to the binary sequence of empty and filled intervals
read from \( x_{2n} \) to \( 0 \). Thus, the Fermi state with a single filled
interval \( [x_{2}=1,\; x_{1}=0] \) reads as unity, the state with a void at
origin: (filled \( [x_{2},\; x_{1}] \) and empty \( [x_{1},\; 0] \)) reads
as \( (10)=2 \), the state with a single gap: \( (101)=3 \), etc. Note that
all even phases have a void at the origin and odd phases have not. 

For free fermions \( V(x)=0 \), \( y(x)=x^{2} \), eqs. (\ref{level}) only
yield the trivial solution corresponding to the Fermi state \( 1 \). In order
to pass to non-trivial realizations of TT, we choose \( U(p) \) corresponding
to the common screened Coulomb potential: 
\begin{equation}
\label{utilde}
U(p)=\frac{4\pi e^{2}}{p^{2}+p_{0}^{2}}.
\end{equation}
 The related explicit form for the kernel 
\begin{equation}
\label{uxx}
u(x,x^{\prime })=\alpha \ln \frac{(x+x^{\prime })^{2}+x_{0}^{2}}{(x-x^{\prime })^{2}+x_{0}^{2}},
\end{equation}
 with the dimensionless screening parameter \( x_{0}=p_{0}/p_{{\textrm{F}}} \)
and the coupling constant \( \alpha =2Me^{2}/\pi p_{{\textrm{F}}} \), evidently
displays the necessary analytical properties for existence of TT. 

With use of eqs. (\ref{vm}) and (\ref{uxx}), the potential \( V\left( x\right)  \)
is expressed in elementary functions:

\begin{equation}
\label{disp2}
V(x)=\sum _{i=1}^{n}V(x;x_{2i-1},x_{2i}),
\end{equation}

\[
V(x;x^{\prime },x^{\prime \prime })=\alpha \left\{ 2\left[ x^{\prime \prime }-x^{\prime }-x_{0}\arctan \frac{2x_{0}\left( x^{\prime \prime }-x^{\prime }\right) \left( x^{2}+x_{0}^{2}+x^{\prime }x^{\prime \prime }\right) }{\left( x^{2}+x_{0}^{2}-x^{\prime 2}\right) \left( x^{2}+x_{0}^{2}-x^{\prime \prime 2}\right) }\right] \right. +\]

\[
\left. +\frac{1}{2x}\left[ \left( x^{\prime \prime 2}+x_{0}^{2}-x^{2}\right) \ln \frac{\left( x+x^{\prime \prime }\right) ^{2}+x_{0}^{2}}{\left( x-x^{\prime \prime }\right) ^{2}+x_{0}^{2}}-\left( x^{\prime 2}+x_{0}^{2}-x^{2}\right) \ln \frac{\left( x+x^{\prime }\right) ^{2}+x_{0}^{2}}{\left( x-x^{\prime }\right) ^{2}+x_{0}^{2}}\right] \right\} .\]
 Then, the straightforward analysis of eqs. (\ref{level}) shows that their
non-trivial solutions only appear when the coupling parameter \( \alpha  \)
exceeds a certain critical value \( \alpha ^{*} \). This corresponds to the
moment when the stability criterion \cite{khosha} \( \sigma (x)=(y_{{\textrm{F}}}(1)-y_{{\textrm{F}}}(x))/(1-x^{2})>0 \)
calculated with the Fermi distribution, \( y_{{\textrm{F}}}(x)=x^{2}+V(x;0,1) \),
fails in a certain point \( 0\leq x_{i}<1 \) within the Fermi sphere: \( \sigma (x_{i})\to 0 \).
There are two different types of such instability depending on the screening
parameter \( x_{0} \) (Fig. 2). For \( x_{0} \) below certain threshold value
\( x_{th}\approx 0.32365 \) (weak screening regime, WSR) the instability point
\( x_{i} \) sets rather close to the Fermi surface: \( 1-x_{i}\ll 1 \), while
it drops in a critical way to zero at \( x_{0}\to x_{th} \) and pertains zero
for all \( x_{0}>x_{th} \) (strong screening regime, SSR). The critical coupling
\( \alpha ^{*}(x_{0}) \) results a monotonically growing function of \( x_{0} \),
having the asymptotics \( \alpha ^{*}(x_{0})\approx \left( \ln 2/x_{0}-1\right) ^{-1} \)at
\( x_{0}\to 0 \) and staying analytic at \( \alpha _{th}=\alpha ^{*}(x_{th})\approx 0.91535 \),
where it only exhibits an inflexion point.\begin{figure}
{\centering \resizebox*{10cm}{!}{\includegraphics{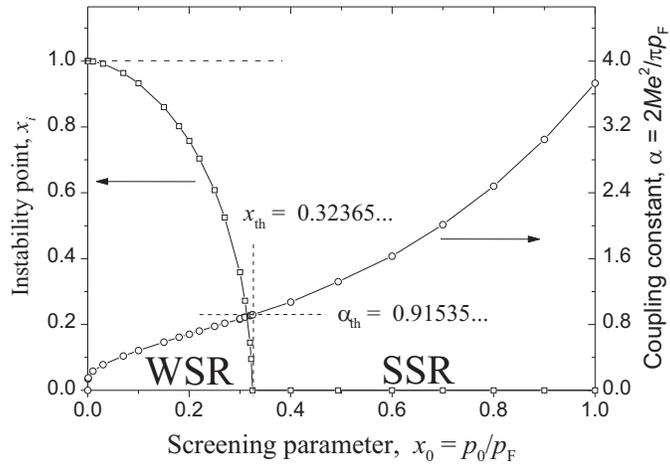}} \par}

\caption{\label{fig2}Critical coupling \protect\( \alpha ^{*}\protect \) and instability
point \protect\( x_{i}\protect \) as functions of screening. The regions of
weak screening (WSR) and strong screening (SSR) are separated by the threshold
value \protect\( x_{th}\protect \) (note that \protect\( x_{th},\, \alpha _{th}\protect \)
is the triple point between the phases 1, 2, 3 in Fig. 3).}
\end{figure} 

These two types of instability give rise to different types of TT from the state
\( 1 \) at \( \alpha >\alpha ^{*} \): at SSR a void appears around \( x=0 \)
(\( 1\to 2 \) transition), and at WSR a gap opens around \( x_{i} \) (\( 1\to 3 \)
transition). Further analysis of eqs. (\ref{level}) shows that the point \( x_{th},\; \alpha _{th} \)
represents a triple point in the phase diagram in variables \( x_{0},\; \alpha  \)
(Fig. 3) where the phases \( 1 \), \( 2 \), and \( 3 \) meet. Similarly to
the onset of instability in the Fermi state \( 1 \), each TT to higher order
phases with growing \( \alpha  \) is manifested by that \( \sigma (x) \),
eq. (\ref{stab}), turns zero at some point \( 0\leq x_{i}<x_{2n} \) different
from the existing interfaces. If this occurs at the very origin, \( x_{i}=0 \),
the phase number rises at TT by \( 1 \), corresponding to opening of a void
(passing from odd to even phase) or to emerging ''island'' (even \( \to  \)
odd). For \( x_{i}>0 \), a thin spherical gap opens within a filled region
or a thin filled spherical sheet emerges within a gap, then the phase number
rises by \( 2 \), not changing the parity. A part of the whole diagram shown
in Fig. 3 demonstrates that with decreasing \( x_{0} \) (weaken screening)
all the even phases terminate at certain triple points. This agrees in particular
with the numeric study of the considered model along the line \( x_{0}=0.07 \)
at growing \( \alpha  \) \cite{zb}, where only the sequence of odd phases
\( 1\to 3\to 5\to \ldots  \) was indicated (shown by the dashed arrow in Fig.
3).\begin{figure}
{\centering \resizebox*{10cm}{!}{\includegraphics{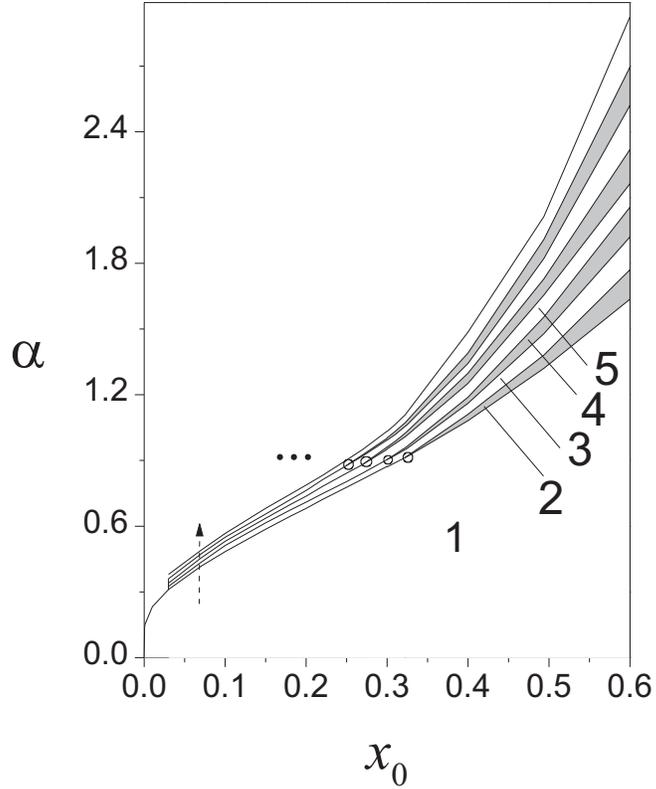}} \par}

\caption{\label{fig3}Phase diagram in variables ``screening-coupling''. Each phase
with certain topology is labeled by the total number of filled and empty regions
(see Fig. \ref{fig1}). Even phases (shadowed) are separated from odd ones by
``\protect\( ^{5}/_{2}\protect \)-kind'' topological transition (TT) lines,
while 2-kind TT lines separate odd phases. Triple points, where two \protect\( ^{5}/_{2}\protect \)-TT
and one 2-TT meet, are shown by circles. The dashed arrow indicates the phase
trajectory studied numerically in Ref. \cite{zb}.}
\end{figure} 

The energy gain \( \Delta \left( \tau \right)  \) at TT as a function of small
overcriticality parameter \( \tau =\alpha /\alpha ^{*}-1 \) is evidently proportional
to \( \tau  \) times the volume of a new emerging phase region (empty or filled).
The analysis shows that the radius of a void (or an island) for a TT with change
of parity is \( \sim \tau ^{1/2} \). Consequently we have \( \Delta \left( \tau \right) \sim \tau ^{5/2} \)
which indicates a similarity of this situation to the known ''\( ^{5}/_{2} \)-kind''
phase transitions in the theory of metals \cite{lifsh}, but its specific character
is that the new segment of Fermi surface opens at very small momentum values,
which can dramatically change the system response to, e.g., electron-phonon
interaction. On the other hand, this segment may have a pronounced effect on
the thermodynamical properties of \( ^{3}{\textrm{He}} \) at low temperatures,
especially in the case of \( P \)-pairing. 

For a TT with unchanged parity, the width of a gap (or a sheet) is found \( \sim \tau  \),
hence the energy gain results \( \Delta \left( \tau \right) \sim \tau ^{2} \),
and such TT can be assigned 2 kind. It follows from the above consideration
that each triple point in the \( x_{0}-\alpha  \) phase diagram is a point
of confluence of two \( ^{5}/_{2} \)-kind TT lines into one 2-kind line. The
latter type of TT was already discussed in literature \cite{llvp,zb}, and we
only mention here that its ocurrance on a whole continuous surface in the momentum
space is rather specific for systems with strong fermion-fermion interaction,
while the known TT's in metals, under the effects of crystalline field, occur
typically at separate points in the quasimomentum space. 

It is of interest to note that in the limit \( x_{0}\to 0,\alpha \to 0 \),
reached along a line \( \alpha =kx_{0} \), we attain the exactly solvable model:
\( U(p)\to (2\pi )^{3}U_{0}\delta (p) \) with \( U_{0}=k/(2Mp_{{\textrm{F}}}) \),
which is known to display FC at all \( U_{0}>0 \) \cite{khosha}. The analytic
mechanism of this behavior consists in that the poles of \( U(p) \), eq. (\ref{utilde})
tend to zero, thus restoring the analytical properties necessary for FC. Otherwise,
the FC regime corresponds to the phase order \( \to \infty  \), when the density
of infinitely thin filled (separated by empty) regions approaches some continuous
function \( 0<\nu (x)<1 \) \cite{zb} and the dispersion law turns flat by
eq. (\ref{level}). 

A few remarks are in order at this point. First, the considered model formally
treats \( x_{0} \) and \( \alpha  \) as independent parameters, though in
fact a certain relation between them can be imposed. Under such restriction,
the system ground state should depend on a single parameter, say the particle
density \( \rho  \), along a certain trajectory \( \alpha (x_{0}) \) in the
above suggested phase diagram. For instance, with the simplest Thomas-Fermi
relation for the free electron gas: \( \alpha (x_{0})=x_{0}^{2}/2 \), this
trajectory stays fully within the Fermi state \( 1 \) over all the physically
reasonable range of densities. Hence a faster growth of \( \alpha (x_{0}) \)
is necessary for realization of TT in any fermionic system with the interaction,
eq. (\ref{utilde}). 

Second, the single particle potential \( V(p) \) of a real system cannot be
an entire function of \( p \) around \( p_{{\textrm{F}}} \) because of the
stepwise form of the quasiparticle distribution. Therefore, as the coupling
constant moves away from the critical value \( \alpha ^{*} \) within the WSR
domain, the concentric Fermi spheres will be taken up by FC. A close look at
the role of the density wave instability, which sets in at sufficiently large
\( \alpha  \), shows that this is true \cite{ksz}. In fact, these arguments
do not work in the case of SSR. Thus, it is quite possible to observe the two
separate Fermi sphere regimes. There is a good reason to mention that neither
in the case when the FC phase transition takes place nor in the case when types
of TT are present the standard Kohn-Sham scheme \cite{wks} is no longer valid.
Beyond the FC or TT phase transitions the occupations numbers of quasiparticles
serve as variational parameters. Thus, to get a reasonable description of the
system, one has to consider the ground state energy as a functional of the occupation
numbers \( E[(n(p)] \) rather then a functional of the density \( E[\rho ] \)
\cite{vs}. A more detailed study of such systems, including the finite temperature
effects, is in order. 

We thank V.A. Khodel and G.E. Volovik for valuable discussions. 

This research was supported in part by the Portuguese program PRAXIS XXI through
the project 2/2.1/FIS/302/94 and under Grant BPD 14226/97 and in part by the
Russian Foundation for Basic Research under Grant 98-02-16170.

\end{document}